\documentclass[cameraready]{Interspeech}

\usepackage{times}
\usepackage{latexsym}
\usepackage[T1]{fontenc}
\usepackage[utf8]{inputenc}
\usepackage{microtype}
\usepackage{inconsolata}
\usepackage{graphicx}
\usepackage{multirow}
\usepackage{booktabs}
\usepackage{amsmath}
\usepackage{makecell}
\usepackage{amsfonts}
\usepackage{utfsym}
\usepackage{arydshln}
\usepackage[most]{tcolorbox}
\usepackage{xcolor} 
\usepackage{float}
\usepackage{enumitem}
\usepackage{pifont}

\usepackage{tikz}
\usetikzlibrary{mindmap}
\usepackage{smartdiagram}
\usesmartdiagramlibrary{additions}
\usepackage{forest}
\usetikzlibrary{shadows}
\usepackage{relsize}

\usepackage{color}
\definecolor{lightcoral}{rgb}{0.94, 0.5, 0.5}
\definecolor{lightgreen}{rgb}{0.56, 0.93, 0.56}
\definecolor{lightyellow}{rgb}{0.94, 0.84, 0.6}
\definecolor{brightlavender}{rgb}{0.75, 0.58, 0.89}
\definecolor{skyblue}{rgb}{0.53, 0.81, 0.92}
\definecolor{peachpuff}{rgb}{1.0, 0.85, 0.73}
\definecolor{goldenrod}{rgb}{0.85, 0.65, 0.13}
\definecolor{orchid}{rgb}{0.85, 0.44, 0.84}
\definecolor{salmon}{rgb}{0.98, 0.5, 0.45}
\definecolor{turquoise}{rgb}{0.25, 0.88, 0.82}
\definecolor{plum}{rgb}{0.87, 0.63, 0.87}
\definecolor{khaki}{rgb}{0.94, 0.9, 0.55}
\definecolor{slateblue}{rgb}{0.42, 0.35, 0.8}
\definecolor{forestgreen}{rgb}{0.13, 0.55, 0.13}
\definecolor{midnightblue}{rgb}{0.1, 0.1, 0.44}
\definecolor{lightsteelblue}{rgb}{0.69, 0.77, 0.87}
\definecolor{limegreen}{rgb}{0.2, 0.8, 0.2}
\definecolor{palegreen}{rgb}{0.6, 0.98, 0.6}
\definecolor{springgreen}{rgb}{0.0, 1.0, 0.5}
\definecolor{mediumseagreen}{rgb}{0.24, 0.7, 0.44}
\definecolor{seagreen}{rgb}{0.18, 0.55, 0.34}
\definecolor{yellowgreen}{rgb}{0.6, 0.8, 0.2}
\definecolor{olivedrab}{rgb}{0.42, 0.56, 0.14}
\definecolor{darkseagreen}{rgb}{0.56, 0.74, 0.56}
\definecolor{lightseagreen}{rgb}{0.13, 0.7, 0.67}
\definecolor{forestgreen}{rgb}{0.13, 0.55, 0.13}
\definecolor{darkolivegreen}{rgb}{0.33, 0.42, 0.18}
\definecolor{greenyellow}{rgb}{0.68, 1.0, 0.18}
\definecolor{chartreuse}{rgb}{0.5, 1.0, 0.0}
\definecolor{mintgreen}{rgb}{0.6, 1.0, 0.6}

\title{SoulX-Duplug: Plug-and-Play Streaming State Prediction Module for \\Realtime Full-Duplex Speech Conversation}

\author[affiliation={1,2,*}]{Ruiqi}{Yan}
\author[affiliation={1,3}]{Wenxi}{Chen}
\author[affiliation={1,3}]{Zhanxun}{Liu}
\author[affiliation={1}]{Ziyang}{Ma}
\author[affiliation={2}]{Haopeng}{Lin}
\author[affiliation={2}]{Hanlin}{Wen}
\author[affiliation={2,4}]{Hanke}{Xie}
\author[affiliation={2}]{Jun}{Wu}
\author[affiliation={1,3}]{Yuzhe}{Liang}
\author[affiliation={1}]{Yuxiang}{Zhao}
\author[affiliation={1}]{Pengchao}{Feng}
\author[affiliation={2}]{Jiale}{Qian}
\author[affiliation={2}]{Hao}{Meng}
\author[affiliation={2,4}]{Yuhang}{Dai}
\author[affiliation={2}]{Shunshun}{Yin}
\author[affiliation={2}]{Ming}{Tao}
\author[affiliation={4}]{Lei}{Xie}
\author[affiliation={1}]{Kai}{Yu}
\author[affiliation={2}, correspondingauthor]{Xinsheng}{Wang}
\author[affiliation={1,3}, correspondingauthor]{Xie}{Chen}

\usepackage{lipsum}
\newcommand\blfootnote[1]{%
  \begingroup
  \renewcommand\thefootnote{}\footnote{#1}%
  \addtocounter{footnote}{-1}%
  \endgroup
}

\address{
    $^1$ X-LANCE Lab, Shanghai Jiao Tong University, China \\
    $^2$ Soul AI Lab, China \\
    $^3$ Shanghai Innovation Institute, China \\
    $^4$ ASLP@NPU, Northwestern Polytechnical University, China
}

\email{yanruiqi@sjtu.edu.cn, wangxinsheng@soulapp.cn, chenxie95@sjtu.edu.cn}


\usepackage{comment}

\begin{document}

\maketitle

\blfootnote{\hspace*{-0.3em}$^*$Work done during an internship at Soul AI Lab.}

\begin{abstract}
Recent advances in spoken dialogue systems have brought increased attention to human-like full-duplex voice interactions. However, our comprehensive review of this field reveals several challenges, including the difficulty in obtaining training data, catastrophic forgetting, and limited scalability. In this work, we propose SoulX-Duplug, a plug-and-play streaming state prediction module for full-duplex spoken dialogue systems. By jointly performing streaming ASR, SoulX-Duplug explicitly leverages textual information to identify user intent, effectively serving as a semantic VAD. To promote fair evaluation, we introduce SoulX-Duplug-Eval, extending widely used benchmarks with improved bilingual coverage. Experimental results show that SoulX-Duplug enables low-latency streaming dialogue state control, and the system built upon it outperforms existing full-duplex models in overall turn management and latency performance. 
We have open-sourced SoulX-Duplug and SoulX-Duplug-Eval\footnote{\scriptsize\url{https://github.com/Soul-AILab/SoulX-Duplug}}.
\end{abstract}

\section{Introduction}

Spoken dialogue models (SDMs) traditionally operate in a half-duplex manner, where listening and speaking are separated in time. In contrast, full-duplex spoken dialogue systems (FD-SDSs)~\cite{defossez2024moshi, wang2024freeze-omni, yu2025salmonn-omni, chen2025minmo, chen2025fireredchat, team2025fun-audio-chat} enable simultaneous listening and speaking, enabling interruption handling, pause handling, and dynamic turn-taking that are essential for more natural interaction. In this work, we propose SoulX-Duplug, a plug-in modeling framework that equips half-duplex systems with full-duplex interaction capability without modifying their backbone architectures.

While recent end-to-end FD-SDSs~\cite{defossez2024moshi, roy2026personaplex, wang2025ntpp} attempt to jointly learn speech understanding, turn-taking, and response generation within a unified framework, such approaches inherently entangle turn-taking policy with language modeling. This tight coupling makes it difficult to disentangle interaction control from content generation, limiting controllability and interpretability. Moreover, the scarcity of large-scale full-duplex conversational data further constrains model generalization and scaling.

\begin{figure}[t]
    \centering
    \includegraphics[width=0.4\textwidth]{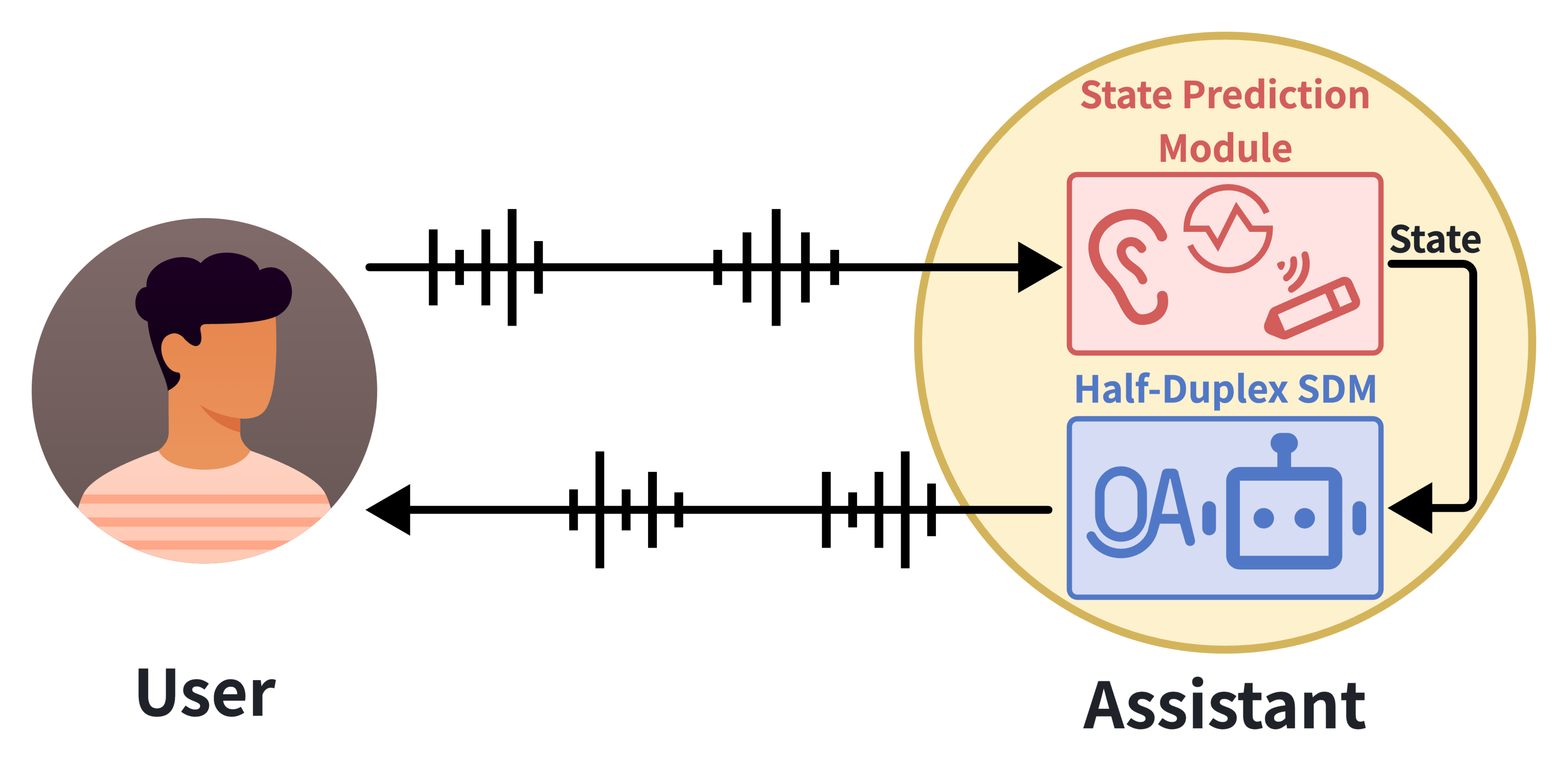}
    \vspace{-2.5mm}
    \caption{Overview of state-driven modular full-duplex speech interaction: the state prediction module processes incoming audio and predicts dialogue states, while the half-duplex SDM generates or stops speech accordingly.}
    \label{fig:overview}
    \vspace{-5.5mm}
\end{figure}

An alternative direction is to equip existing half-duplex spoken dialogue models with external full-duplex capabilities, commonly realized as a “VAD–ASR–Turn Detection” module, thereby forming modular full-duplex systems \cite{chen2025fireredchat, zhang2025Tencent-modular-system} (Figure~\ref{fig:overview}).
Most existing implementations adopt cascaded pipelines for dialogue state control. In such architectures, a VAD module first detects and segments active speech from the continuous audio stream, which is then passed to an ASR module for transcription. Based on the recognized text, a turn detection model determines whether the half-duplex SDM should respond. However, conventional VAD methods primarily rely on acoustic features and lack access to semantic information. Moreover, non-streaming ASR and turn-detection components introduce latency that grows with input length, thereby reducing responsiveness in real-time interaction. Although recent efforts have introduced end-to-end semantic-aware VAD variants \cite{wu2025phoenix-vad, liao2025flexduo}, these approaches generally do not explicitly leverage textual representations as inputs, leaving the integration between speech recognition and semantic turn detection inadequately explored.

With respect to evaluation, existing work on FD-SDSs largely relies on self-constructed test sets. The lack of a widely accepted benchmark hinders fair comparison and makes it difficult to quantify progress across different approaches. In addition, publicly available benchmarks that support bilingual or multilingual evaluation remain scarce, further constraining cross-lingual analysis and broader generalization assessment.

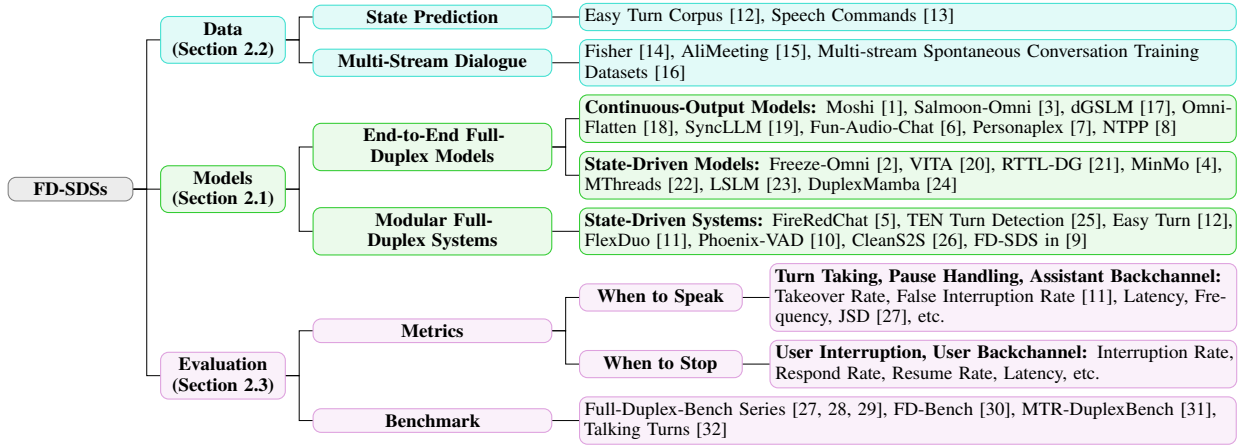
\begin{figure*}[ht]
\centering
  \tikzset{
          my node/.style={
              draw,
              align=center,
              thin,
              text width=1.2cm, 
              rounded corners=3,
          },
          my leaf/.style={
              draw,
              align=left,
              thin,
              text width=8.5cm, 
              rounded corners=3,
          }
  }
  \forestset{
    every leaf node/.style={
      if n children=0{#1}{}
    },
    every tree node/.style={
      if n children=0{minimum width=1em}{#1}
    },
  }
  \begin{forest}
      nonleaf/.style={font=\bfseries\scriptsize},
       for tree={%
          every leaf node={my leaf, font=\scriptsize},
          every tree node={my node, font=\scriptsize, l sep-=4.5pt, l-=1.pt},
          anchor=west,
          inner sep=2pt,
          l sep=10pt, 
          s sep=3pt, 
          fit=tight,
          grow'=east,
          edge={ultra thin},
          parent anchor=east,
          child anchor=west,
          if n children=0{}{nonleaf}, 
          edge path={
              \noexpand\path [draw, \forestoption{edge}] (!u.parent anchor) -- +(5pt,0) |- (.child anchor)\forestoption{edge label};
          },
          if={isodd(n_children())}{
              for children={
                  if={equal(n,(n_children("!u")+1)/2)}{calign with current}{}
              }
          }{}
      }
[FD-SDSs, draw=gray, fill=gray!15, text width=1.5cm, text=black
    [{Data \\ (Section~\ref{sec:related_data})}, color=turquoise, fill=turquoise!15, text width=1.5cm, text=black
        [State Prediction, color=turquoise, fill=turquoise!15, text width=3cm, text=black
            [{Easy Turn Corpus \cite{li2025easy-turn}, Speech Commands \cite{warden2018speech-commands}}, color=turquoise, fill=turquoise!15, text width=8.5cm, text=black]
        ],
        [Multi-Stream Dialogue, color=turquoise, fill=turquoise!15, text width=3cm, text=black
            [{Fisher \cite{cieri2004fisher}, AliMeeting \cite{yu2022alimeeting}, Multi-stream Spontaneous Conversation Training Datasets \cite{zhou2025multistream-dataset}}, color=turquoise, fill=turquoise!15, text width=8.5cm, text=black]
        ],
    ],
    [{Models \\ (Section~\ref{sec:related_models})}, color=limegreen, fill=limegreen!10, text width=1.5cm, text=black
        [{End-to-End Full-Duplex Models}, color=limegreen, fill=limegreen!10, text width=3cm, text=black
            [{\textbf{Continuous-Output Models:} Moshi \cite{defossez2024moshi}, Salmoon-Omni \cite{yu2025salmonn-omni}, dGSLM \cite{nguyen2023dGSLM}, OmniFlatten \cite{zhang2024omniflatten}, SyncLLM \cite{veluri2024SyncLLM}, Fun-Audio-Chat \cite{team2025fun-audio-chat}, Personaplex \cite{roy2026personaplex}, NTPP \cite{wang2025ntpp}}, color=limegreen, fill=limegreen!10, text width=8.5cm, text=black],
            [{\textbf{State-Driven Models:} Freeze-Omni \cite{wang2024freeze-omni}, VITA \cite{fu2024vita}, RTTL-DG \cite{mai2025RTTL-DG}, MinMo \cite{chen2025minmo}, MThreads \cite{wang2024MThreads}, LSLM \cite{ma2025lslm}, DuplexMamba \cite{lu2025duplexmamba}}, color=limegreen, fill=limegreen!10, text width=8.5cm, text=black]
        ],
        [Modular Full-Duplex Systems, color=limegreen, fill=limegreen!10, text width=3cm, text=black
            [{\textbf{State-Driven Systems:} FireRedChat \cite{chen2025fireredchat}, TEN Turn Detection \cite{TEN_Turn_Detection}, Easy Turn \cite{li2025easy-turn}, FlexDuo \cite{liao2025flexduo}, Phoenix-VAD \cite{wu2025phoenix-vad}, CleanS2S \cite{lu2025cleans2s}, FD-SDS in \cite{zhang2025Tencent-modular-system}}, color=limegreen, fill=limegreen!10, text width=8.5cm, text=black]
        ],
    ],
    [{Evaluation \\ (Section~\ref{sec:related_eval})}, color=plum, fill=plum!15, text width=1.5cm, text=black
        [{Metrics}, color=plum, fill=plum!15, text width=3cm, text=black
            [{When to Speak}, color=plum, fill=plum!15, text width=2cm, text=black
                [{\textbf{Turn Taking, Pause Handling, Assistant Backchannel:} Takeover Rate, False Interruption Rate \cite{liao2025flexduo}, Latency, Frequency, JSD \cite{lin2025full-duplex-bench1}, etc.}, color=plum, fill=plum!15, text width=6cm, text=black],
            ],
            [{When to Stop}, color=plum, fill=plum!15, text width=2cm, text=black
                [{\textbf{User Interruption, User Backchannel:} Interruption Rate, Respond Rate, Resume Rate, Latency, etc.}, color=plum, fill=plum!15, text width=6cm, text=black],
            ],
        ],
        [Benchmark, color=plum, fill=plum!15, text width=3cm, text=black
            [{Full-Duplex-Bench Series \cite{lin2025full-duplex-bench1, lin2025full-duplex-bench1.5, lin2025full-duplex-bench2}, FD-Bench \cite{peng2025fd-bench}, MTR-DuplexBench \cite{zhang2025MTR-DuplexBench}, Talking Turns \cite{arora2025talking-turns}}, color=plum, fill=plum!15, text width=8.5cm, text=black]
        ],
    ],
]
    \end{forest}
  \caption{Overview of Researches on Full-Duplex Spoken Dialogue Systems.}
  \vspace{-3mm}
  \label{fig:related}
\end{figure*}


To fill the gaps, we propose \textbf{SoulX-Duplug}, a unified streaming state prediction framework for modular full-duplex spoken dialogue systems, together with a bilingual evaluation benchmark, \textbf{SoulX-Duplug-Eval}. 
We formulate duplex interaction control as a streaming state prediction problem under incrementally available observations. Rather than decomposing VAD, ASR, and turn detection into independent cascaded modules, SoulX-Duplug models them within a single streaming architecture. A key design principle is to incorporate semantic supervision through a streaming ASR objective during training, enabling the model to learn linguistically informed representations that enhance state prediction.
By performing ASR and state prediction jointly in a chunk-based streaming manner, the model leverages text information for semantic-level turn detection while maintaining low latency, functioning effectively as a semantic VAD module.
To facilitate standardized evaluation, we further present SoulX-Duplug-Eval, a bilingual benchmark suite for both state-level and system-level assessment.
For duplex state prediction, it extends the Easy Turn testset \cite{li2025easy-turn} by incorporating additional English samples. For system-level assessment of FD-SDSs, it selects representative tasks from the Full-Duplex-Bench series \cite{lin2025full-duplex-bench1, lin2025full-duplex-bench1.5, lin2025full-duplex-bench2} and further supplements them with Chinese test sets to enhance bilingual coverage. We expect that these benchmarks will facilitate more comparable and standardized bilingual evaluation in full-duplex spoken dialogue research.

Experimental results on Full-Duplex-Bench and SoulX-Duplug-Eval indicate that SoulX-Duplug achieves low latency close to its theoretical bound, and the full-duplex spoken dialogue system built upon it attains the best overall performance in terms of turn management and latency across multiple evaluation dimensions, confirming its practical utility in downstream applications. Further analysis on Easy Turn testset demonstrates that SoulX-Duplug effectively leverages auxiliary textual supervision to improve dialogue state prediction, validating the proposed text-guided streaming state prediction design.
We promise to open-source SoulX-Duplug and SoulX-Duplug-Eval for future development in this field.

In conclusion, our main contributions include:
\begin{itemize}
    \item We propose SoulX-Duplug, a bilingual streaming state prediction module that unifies VAD, ASR, and turn detection within a single framework. By incorporating text-guided streaming state prediction, it enables low-latency and semantically informed dialogue state control for modular full-duplex spoken dialogue systems.
    \item We release SoulX-Duplug-Eval, a complementary benchmark that extends the Easy Turn testset and Full-Duplex-Bench, promoting more standardized and comparable cross-lingual assessment in future research.
    \item SoulX-Duplug enables streaming turn detection with an average latency of 240\,ms. A FD-SDS built upon SoulX-Duplug outperforms existing systems across multiple evaluation settings, validating its downstream practicality. Further experiments verify the effectiveness of our proposed methods.
\end{itemize}

\section{Related Work}
\label{sec:review}

\begin{figure*}[t]
    \centering
    \includegraphics[width=0.9\textwidth]{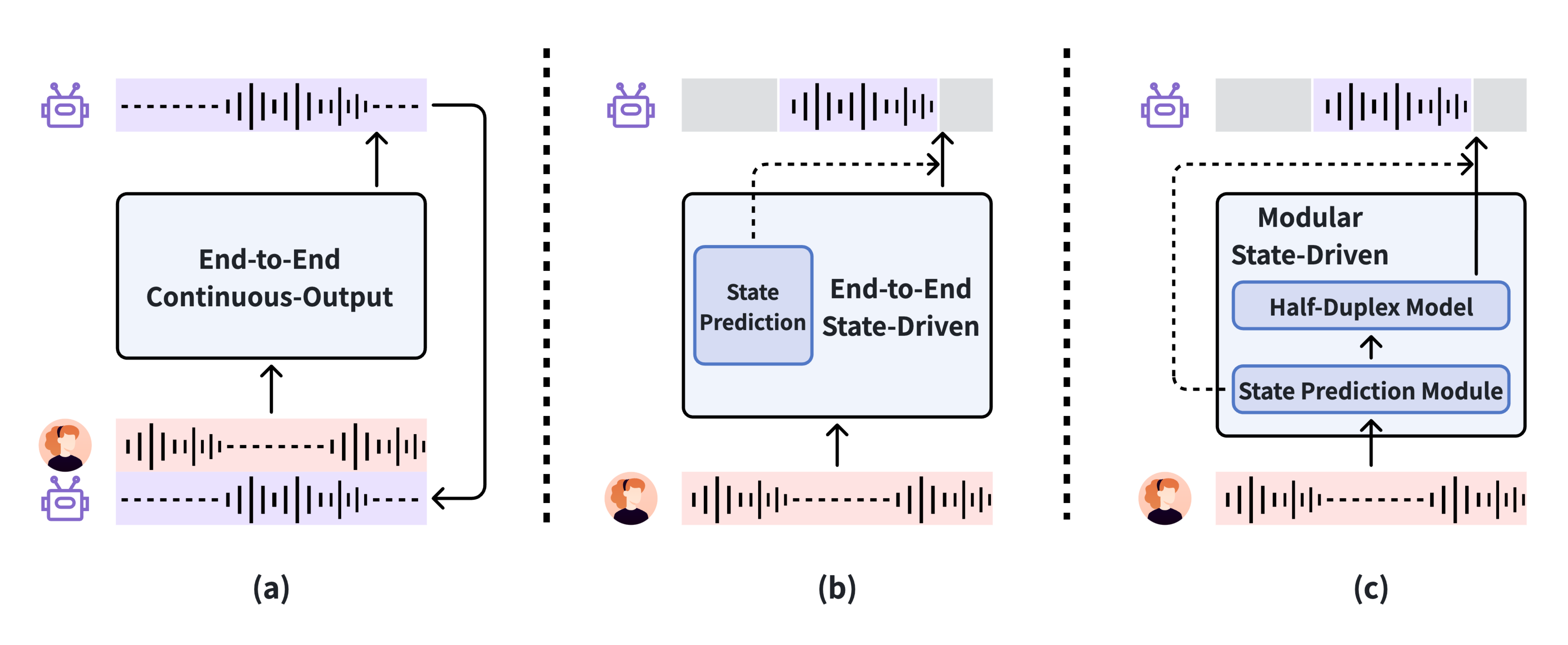}
    \vspace{-6.5mm}
    \caption{Illustration of 3 types of full-duplex spoken dialogue systems. (a): End-to-End Continuous-Output Full-Duplex Models. (b): End-to-End State-Driven Full-Duplex Models. (c): Modular State-Driven Full-Duplex Systems.}
    \label{fig:types}
    \vspace{-3mm}
\end{figure*}

\subsection{Full-Duplex SDMs}
\label{sec:related_models}
From a modeling perspective, we divide current FD-SDSs into two categories: end-to-end models and modular systems, as presented in Figure~\ref{fig:related} and Figure~\ref{fig:types}.
End-to-end approaches offer the conceptual advantage of jointly modeling speech understanding and generation within a unified framework. End-to-end continuous-output models~\cite{defossez2024moshi, roy2026personaplex, yu2025salmonn-omni} adopt a dual-stream formulation in which the model jointly processes the user-side input stream and the model-side output stream as parallel inputs. At each time step, the model conditions on both the incoming user speech and its own previously generated responses, and directly predicts the next segment of the model output stream. This design enables unified modeling of overlapping speech and response generation within a single framework.
End-to-end state-driven models separate response generation from dialogue state control. Models like Freeze-Omni \cite{wang2024freeze-omni} and MinMo \cite{chen2025minmo}, continuously process the user’s speech input while an internal predictor, typically derived from hidden representations, dynamically estimates the current dialogue state (e.g., listening or speaking). The predicted state then governs whether the model produces a spoken response or remains silent. MThreads \cite{wang2024MThreads}, DuplexMamba \cite{lu2025duplexmamba}, and LSLM \cite{ma2025lslm} predict special state tokens to manage a finite state machine for the dialogue. The state-driven approach introduces an explicit controller within the end-to-end architecture, enabling structured management of turn-taking behavior while retaining integrated modeling of speech understanding and generation.
In practice, however, both continuous-output and state-driven models face substantial challenges. High-quality full-duplex training data with precise temporal alignment are costly to collect and annotate, making supervised fine-tuning difficult and prone to performance degradation or overfitting to specific interaction patterns or domains. Furthermore, end-to-end models require persistent computation during real-time inference, and this always-on processing paradigm poses challenges for scaling up model size considering realistic latency constraints, computational overhead, and deployment cost.

Modular architectures, which combine an independent duplex state predictor with a half-duplex spoken dialogue model, offer improved engineering controllability and deployment flexibility. A lightweight dialogue management module can greatly reduce real-time computational overhead, thereby decoupling latency-sensitive state control from response generation and enabling the half-duplex spoken dialogue model to scale up.
Several representative systems adopt cascaded pipelines for state control. For example, FiredRedChat \cite{chen2025fireredchat} and TEN Turn Detection \cite{TEN_Turn_Detection} employ a sequential combination of VAD, ASR, and text-based turn detection. Easy Turn integrates ASR and turn detection into a unified model, but still depends on an external VAD to segment input audio streams. Although such designs improve modularity, the underlying VAD components are typically based on acoustic features and lack semantic awareness, which limits their ability to identify turn boundaries and interruptions at the semantic level. In addition, the ASR and turn-detection models operate in a non-streaming fashion, introducing additional latency especially for longer inputs.
Other efforts attempt to manage dialogue state in an end-to-end manner. FlexDuo \cite{liao2025flexduo} proposes a decoupled framework for full-duplex dialogue control and generation. However, its backbone model has a size of 7B parameters, which may cause considerable computational pressure for real-time deployment. Phoenix-VAD \cite{wu2025phoenix-vad} introduces an LLM-based approach for streaming semantic endpoint detection. While it enhances semantic awareness in turn boundary detection, it does not incorporate ASR functionality and does not explicitly leverage text as input. As shown in Table~\ref{tab:comparison}, SoulX-Duplug is the first ASR-assisted streaming state prediction module for modular FD-SDSs.

\begin{table}[htbp]
\centering
\setlength{\abovecaptionskip}{.2cm}
\setlength{\belowcaptionskip}{.0cm}
\resizebox{1\linewidth}{!}{
\begin{tabular}{lcccc}
\toprule
\multirow{2}{*}{\textbf{Models}} & \multirow{2}{*}{\textbf{Streaming}} & \multirow{2}{*}{\textbf{ASR}} & \multirow{2}{*}{\makecell{\textbf{End-to-End} \\ \textbf{Optimization}}} & \multirow{2}{*}{\textbf{Language}} \\
\\
\midrule
Easy Turn & \textcolor{red}{\ding{55}} & \textcolor{green!50!black}{\ding{51}} & \textcolor{green!50!black}{\ding{51}} & zh \\
FireRedChat & \textcolor{red}{\ding{55}} & \textcolor{green!50!black}{\ding{51}} & \textcolor{red}{\ding{55}} & en, zh \\
Phoenix-VAD & \textcolor{green!50!black}{\ding{51}} & \textcolor{red}{\ding{55}} & \textcolor{green!50!black}{\ding{51}} & en, zh \\
Flexduo & \textcolor{green!50!black}{\ding{51}} & \textcolor{red}{\ding{55}} & \textcolor{green!50!black}{\ding{51}} & en, zh \\
\textbf{SoulX-Duplug (Ours)} & \textcolor{green!50!black}{\ding{51}} & \textcolor{green!50!black}{\ding{51}} & \textcolor{green!50!black}{\ding{51}} & en, zh \\ 
\bottomrule
\end{tabular}
}
\vspace{2mm}
\caption{Comparison with existing plug-and-play state prediction modules for full-duplex speech conversation.}
\vspace{-6mm}
\label{tab:comparison}
\end{table}

\begin{figure*}[t]
    \centering
    \includegraphics[width=0.9\textwidth]{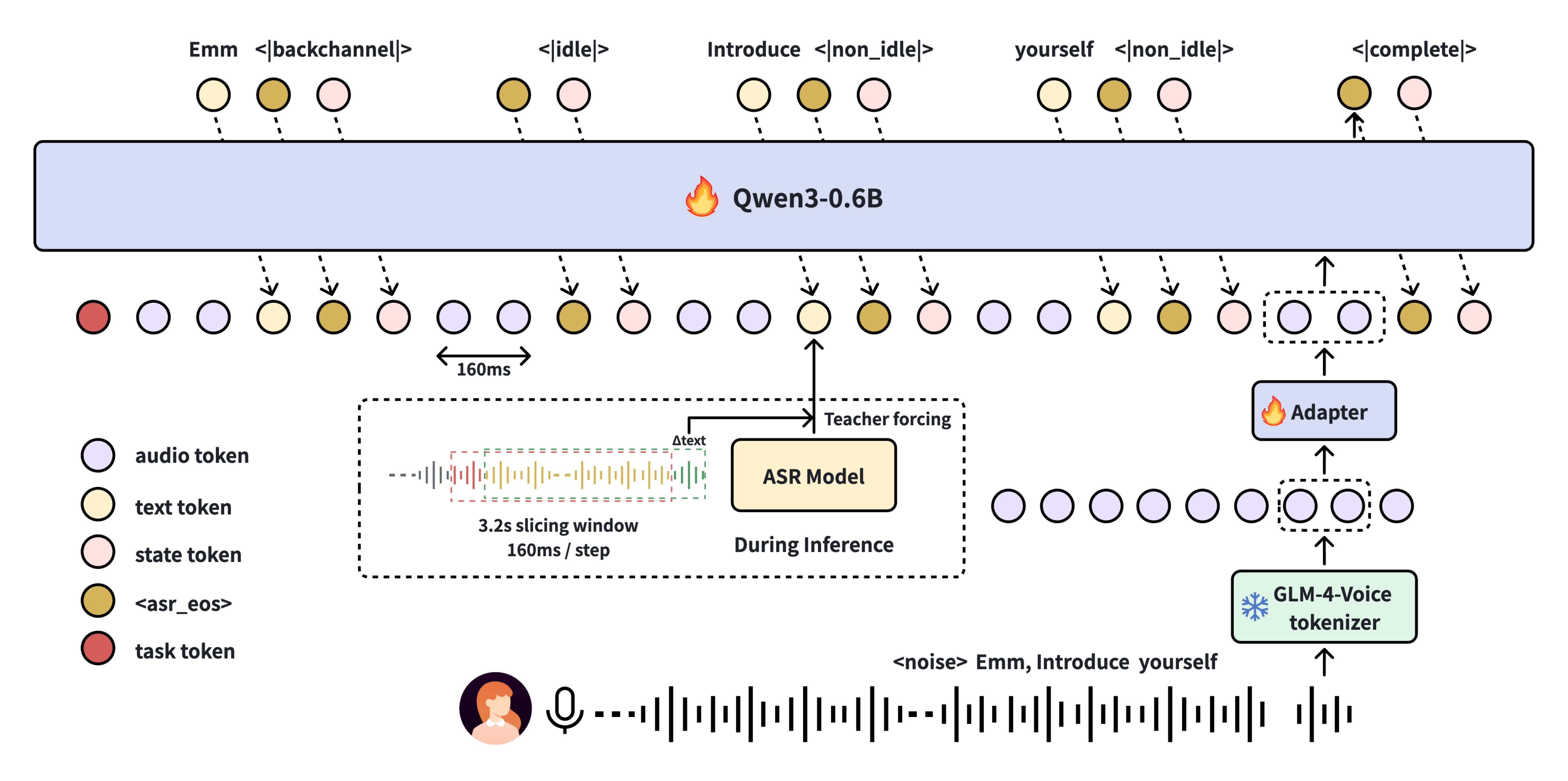}
    \vspace{-4mm}
    \caption{The architecture of SoulX-Duplug. The model runs with interleaved audio tokens, text tokens, and state tokens, ensuring that textual information is available when predicting state tokens. During training, VAD, ASR, and state prediction are end-to-end optimized. During inference, a lightweight state-of-the-art ASR model provides text guidance via teacher forcing.}
    \label{fig:model}
    \vspace{-2mm}
\end{figure*}

\subsection{Data for Full-Duplex SDMs}
\label{sec:related_data}
For the training of state prediction models, Easy Turn Corpus \cite{li2025easy-turn} and Speech Commands \cite{warden2018speech-commands} are speech datasets annotated with dialogue states or user intentions (e.g., semantically complete, semantically incomplete, interruption, backchannel, etc.). These datasets typically provide fine-grained labels describing the role of each single-round utterance.

As for multi-speaker, multi-channel conversational corpora, Fisher \cite{cieri2004fisher} provides large-scale English two-channel telephone conversation samples. AliMeeting \cite{yu2022alimeeting} contains multi-channel recordings of real-world Chinese meeting discussions. The Multi-stream Spontaneous Conversation Training Datasets \cite{zhou2025multistream-dataset} offers open-source dual-stream conversational speech data covering both Chinese and English, with relatively high recording quality. However, its total duration is only 15 hours.
These publicly available resources are either limited in scale or domain diversity, and are concentrated on constrained scenarios such as meetings or casual conversations. Datasets that cover general knowledge, multi-turn reasoning, and complex contextual understanding are particularly scarce and difficult to construct. This data limitation directly constrains the intelligence and generalization ability of full-duplex systems in real-world applications. Our proposed SoulX-Duplug decouples turn taking from dialogue modeling, enabling arbitrary half-duplex systems to acquire full-duplex capability in a plug-and-play manner. This design allows subsequent improvements and scaling in dialogue modeling to rely solely on half-duplex training data that are relatively easier to obtain, thus alleviating the data challenge.

\subsection{Evaluation of Full-Duplex SDMs}
\label{sec:related_eval}
Although existing evaluations of full-duplex spoken dialogue systems differ in terminology, their underlying task formulations and basic evaluation dimensions are largely consistent. Most benchmarks focus on two core tasks. The first concerns “when to speak”, namely, whether the system can take the turn at an appropriate moment after the user finishes speaking. The second concerns “when to stop”, that is, whether the system can timely terminate its output when the user initiates an interruption. To assess these capabilities, three main types of metrics are commonly adopted. Success rate measures the proportion of correct turn-taking or successful interruption handling. Error rate, including false start and false stop rates, quantifies inappropriate responses or premature termination. Latency-related metrics, such as turn-taking delay and stop delay, evaluate the temporal responsiveness of the system. Some benchmarks further extend the evaluation to higher-level interaction behaviors, including assistant backchannel and proactive interruption \cite{lin2025full-duplex-bench1}.

Several benchmarks have been proposed in this area. The Full-Duplex-Bench series \cite{lin2025full-duplex-bench1, lin2025full-duplex-bench1.5, lin2025full-duplex-bench2} covers multiple scenarios, including turn-taking capability, overlap handling, and multi-turn evaluation, but its test sets are limited to English. FD-Bench \cite{peng2025fd-bench} provides a comprehensive benchmarking pipeline for full-duplex spoken dialogue systems, automating dialogue generation, speech corpus simulation, and multi-criteria assessment. MTR-DuplexBench \cite{zhang2025MTR-DuplexBench} introduces a multi-round evaluation protocol using both natural and synthetic dialogue data. For the specific task of duplex state prediction, the Easy Turn testset \cite{li2025easy-turn} is designed to evaluate turn detection performance. However, it currently supports only Chinese evaluation. By introducing SoulX-Duplug-Eval, we aim to enhance the cross-lingual comparability of existing benchmarks.


\section{SoulX-Duplug}

\subsection{Overview}

As shown in Figure~\ref{fig:model}, SoulX-Duplug is a unified streaming module for realtime full-duplex spoken dialogue that integrates VAD, ASR, and state prediction within a single framework. The model uses discrete speech tokens extracted from user-side audio input and streamingly produces interleaved ASR outputs and dialogue state tokens.

\begin{figure*}[t]
    \centering
    \includegraphics[width=\textwidth]{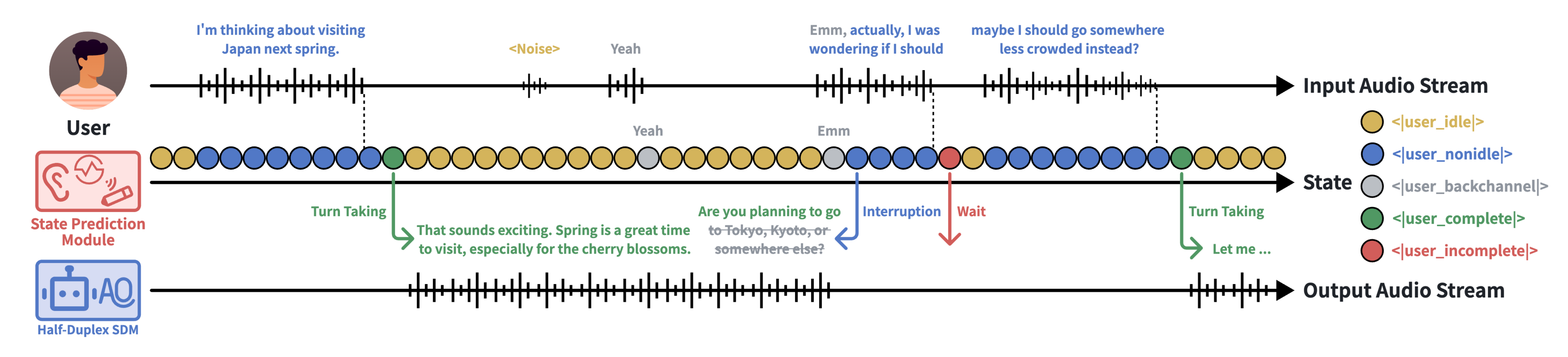}
    \vspace{-6mm}
    \caption{A detailed example explaining the state token design of SoulX-Duplug and its streaming inference paradigm.}
    \label{fig:state-illustration}
    \vspace{-3mm}
\end{figure*}

\subsection{State Token Design}

We define five state tokens to model the interaction dynamics in full-duplex spoken dialogue, with an example illustrated in Figure~\ref{fig:state-illustration}.
\begin{itemize}
    \item \texttt{<|user\_idle|>} indicates that the current audio chunk contains no semantic content, such as silence or noise.
    \item \texttt{<|user\_nonidle|>} denotes that the chunk contains semantically meaningful speech. 
    \item \texttt{<|user\_backchannel|>} represents user backchannel behavior.
    \item \texttt{<|user\_complete|>} indicates that the user’s utterance is semantically complete and the assistant may take the turn. 
    \item \texttt{<|user\_incomplete|>} represents that the user pauses but his/her utterance is semantically incomplete so the assistant should wait.
\end{itemize}

\subsection{Speech Input Modeling}

We adopt the GLM-4-Voice tokenizer \cite{zeng2024glm-4-voice} to extract audio tokens $A_d = [a_{d,1}, a_{d,2}, \dots, a_{d,N}]$ at a frequency of 12.5\,Hz. The tokenizer is a block-causal speech tokenizer pretrained on large-scale speech data and serves as the foundational speech encoder for bilingual speech understanding \cite{zeng2024scaling}. For streaming inference, we use a block size of 12 for audio token generation. For each step, the model encodes a target window of 160\,ms with a left context (look-back) of 960\,ms and a right context (look-ahead) of 40\,ms, resulting in a total receptive field of 1160\,ms and 15 extracted tokens. The tokens corresponding to the target region are aligned with the second-to-last and third-to-last tokens within the block. Then a linear encoder projector transforms the embedding of $A_d$ into feature $A$ to ensure alignment with LLM’s embedding dimension, defined as $A = MLP(A_d)$.

\subsection{Text-Guided Streaming State Prediction}

To explicitly incorporate semantic information into streaming state prediction, we innovatively introduce a joint ASR objective and design an interleaved prediction paradigm:
\begin{equation}
\{A_1, T_1, S_1, A_2, T_2, S_2, \dots, A_T, T_T, S_T\}
\end{equation}

Each 160\,ms audio chunk corresponds to two audio tokens, $A_t = [a_{t,1}, a_{t,2}]$ for $t$-th chunk. 
Conditioned on the historical context $\mathcal{H}_{t-1}$, the model first predicts the ASR token sequence for the current chunk:
\begin{equation}
T_t \sim P(T_t \mid A_{\le t}, T_{<t}, S_{<t})
\end{equation}
where $T_t$ represents the streaming ASR output aligned to chunk $t$. 
After generating $T_t$, the model predicts the dialogue state token:
\begin{equation}
S_t \sim P(S_t \mid A_{\le t}, T_{\le t}, S_{<t})
\end{equation}
$S_t$ denotes the duplex state associated with the current chunk. 
This interleaved design enables explicit semantic guidance for state prediction while maintaining streaming inference.

\subsection{Training Objective}

Since different token types (e.g., text tokens, \texttt{<asr\_eos>}, and various state tokens) occur with significantly different frequencies in long sequences, we adopt a weighted token-level training objective. Let $\mathcal{Y}$ denote the full target token sequence and $y_j$ the $j$-th token. The overall loss is defined as
{
\setlength{\abovedisplayskip}{3pt}
\setlength{\belowdisplayskip}{3pt}
\begin{equation}
\mathcal{L} = \sum_{j=1}^{|\mathcal{Y}|} 
\lambda_{\tau(y_j)} 
\cdot 
\mathcal{L}_{\text{CE}}(y_j)
\end{equation}
}
where $\mathcal{L}_{\text{CE}}(y_j)$ denotes the cross-entropy loss for predicting token $y_j$, $\tau(y_j)$ maps each token to its token type, and $\lambda_{\tau(y_j)}$ is a token-type-specific weighting coefficient used to balance optimization across heterogeneous token categories.

\subsection{Hybrid 3-stage Training with Teacher-Forced Inference}
\begin{figure*}[htbp]
    \centering
    \includegraphics[width=0.9\textwidth]{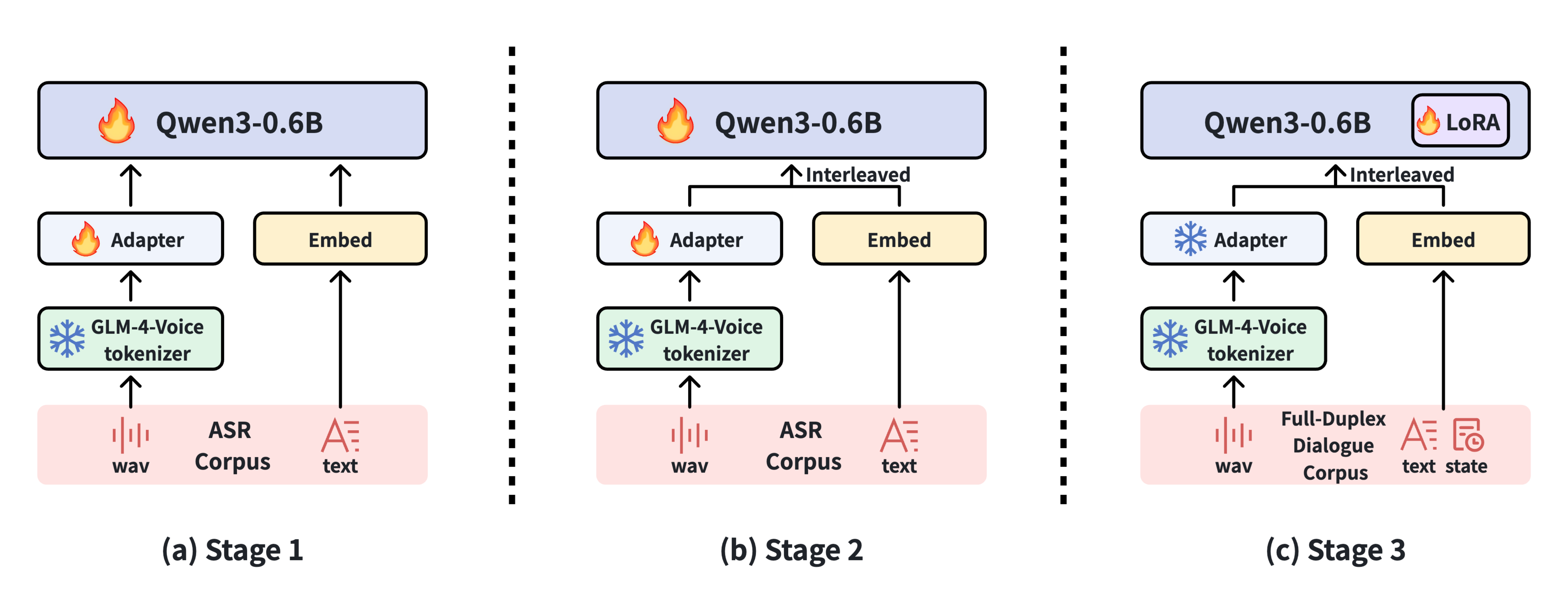}
    \vspace{-5mm}
    \caption{The three-stage training phases of SoulX-Duplug. (a) Stage 1: Non-Streaming ASR Pretraining. (b) Stage 2: Streaming ASR Adaptation. (c) Stage 3: State Prediction SFT.}
    \label{fig:training}
    \vspace{-5mm}
\end{figure*}

As demonstrated in Figure~\ref{fig:training}, we design three sequential training stages: non-streaming ASR pretraining, streaming ASR adaptation, and duplex state prediction fine-tuning. The first two stages focus on speech recognition capability, while the final stage specializes the model for real-time dialogue management.
Moreover, SoulX-Duplug uses a hybrid training–inference strategy. During the third stage training, the model is optimized in an end-to-end manner over VAD, ASR and state prediction tasks. During inference, we employ a state-of-the-art yet lightweight external ASR model (i.e., SenseVoice Small \cite{an2024sensevoice}) to provide teacher-forced streaming ASR outputs for each chunk. This design preserves the benefits of unified end-to-end training while ensuring more accurate and efficient real-time deployment.

\subsection{Algorithmic Latency}
The audio chunk has a fixed duration of 160\,ms. Assume that the user’s speech terminates within chunk $t_i$. Because the model operates in a streaming manner, it can only determine the absence of subsequent active speech after the arrival of the next chunk. When processing chunk $t_{i+1}$, the model detects no active speech, considers the \texttt{<|user\_nonidle|>} state to have ended, and then performs a decision on whether the preceding semantic content is complete or not (i.e., \texttt{<|user\_complete|>} or \texttt{<|user\_incomplete|>}).
Since the actual endpoint of the user’s speech may occur at any position within chunk $t_i$, under a uniform distribution assumption, 
the theoretical average latency of SoulX-Duplug is: $\text{Latency}_{\text{avg.}} = 80\,\text{ms} + 160\,\text{ms} = 240\,\text{ms}.$

\section{SoulX-Duplug-Eval}

To address the lack of cross-lingual evaluation resources in existing full-duplex spoken dialogue benchmarks, we constructed complementary test sets to improve comparability across different models and support standardized and fair comparison in both dialogue state prediction and system-level full-duplex dialogue settings.

\subsection{Bilingual Easy Turn Testset}

We introduce Easy Turn testset-En as the English counterpart of the original Easy Turn testset \cite{li2025easy-turn}. It focuses on duplex state prediction and contains two categories: \textit{Complete} and \textit{Incomplete}. The \textit{Complete} category consists of semantically complete utterances generated by Chat-GPT and synthesized with ChatTTS\footnote{\scriptsize\url{https://github.com/2noise/ChatTTS}} \cite{chattts}, comprising 318 samples. The \textit{Incomplete} category contains semantically incomplete utterances generated in the same manner, with 299 samples.

\subsection{Bilingual Full-Duplex-Bench}

To support system-level evaluation in Chinese, we construct Full-Duplex-Bench-Zh as a Chinese counterpart to Full-Duplex-Bench \cite{lin2025full-duplex-bench1, lin2025full-duplex-bench1.5}. The dataset covers four representative interaction scenarios. All textual content is generated with Chat-GPT and synthesized via state-of-the-art TTS systems.

\begin{itemize}
    \item The \textit{Turn-Taking} subset contains 155 samples, where a user speaks for several seconds followed by 15 seconds of silence. The speech is synthesized using ChatTTS.
    \item The \textit{Pause Handling} subset includes 239 samples, in which a single utterance contains inserted pauses generated via the \texttt{[uv\_break]} control token in ChatTTS.
    \item The \textit{User Backchannel} subset comprises 199 samples, structured as user speech followed by 3 seconds of silence, a short backchannel utterance, and 15 seconds of silence. This subset is synthesized using SoulX-Podcast \cite{xie2025SoulX-Podcast}.
    \item The \textit{User Interruption} subset contains 161 samples, where an initial user utterance is followed by 3 seconds of silence and a semantically related second utterance, then 15 seconds of silence. The speech is synthesized using ChatTTS.
\end{itemize}

Together, these datasets provide cross-lingual coverage for both duplex state prediction and full-duplex dialogue evaluation, facilitating more consistent and comparable experimental analysis.

\section{Experimental Setup}

\subsection{Training Details}

For ASR training of SoulX-Duplug, we use large-scale Mandarin and English corpora. The Mandarin data include \textit{AISHELL-1} \cite{bu2017aishell-1}, \textit{AISHELL-3} \cite{shi2020aishell-3}, \textit{WenetSpeech} \cite{zhang2022wenetspeech}, and the \textit{CommonVoice-CN}, \textit{Emilia-CN}, and \textit{MAGICDATA} subsets from VoxBox \cite{wang2025spark-tts}, totaling approximately 47,000 hours. The English data include \textit{LibriSpeech} \cite{panayotov2015librispeech}, \textit{GigaSpeech} \cite{chen2021gigaspeech}, and the \textit{CommonVoice-EN} and \textit{Emilia-EN} subsets from VoxBox, totaling approximately 31,000 hours. For streaming ASR training, we first obtain character-level or word-level alignments and then reorganize the data in an interleaved chunk-based format. Timestamps of Mandarin corpus are generated using Paraformer\footnote{\scriptsize iic/speech\_paraformer-large-vad-punc\_asr\_nat-zh-cn-16k-common-vocab8404-pytorch}, while timestamps of English corpus are produced using WhisperX \cite{bain2023whisperx}.

In the state prediction training stage, we use the Fisher dataset \cite{cieri2004fisher} for English at the thousand-hour level. For Mandarin, since no suitable open-source dataset is available, we employ a ten-thousand-hour-level in-house corpus constructed in the same format as Fisher.
The annotation pipeline is carefully designed. We first perform alignment, and for Mandarin data, we filter samples based on dual-ASR consistency. To improve robustness, we introduce noise augmentation by adding Musan \cite{snyder2015musan} noise globally and ESC-50 \cite{piczak2015ESC-50} noise to silence segments. State labels are annotated using Qwen2.5-72B-Instruct \cite{Yang2024Qwen2.5}. SoulX-Duplug adopts a pretrained GLM-4-Voice tokenizer \cite{zeng2024glm-4-voice} as the speech encoder and Qwen3-0.6B \cite{yang2025qwen3} as the LLM backbone. The speech tokenizer remains frozen throughout training. During ASR pretraining, both the LLM and adapter layers are fully fine-tuned. During state prediction training, we apply LoRA \cite{hu2022lora} fine-tuning with rank $r=32$ to the LLM on bilingual trainsets. All training is conducted on NVIDIA H20 GPUs.

\subsection{Evaluation Setup}
During inference, teacher-forcing ASR is applied to provide more accurate textual guidance. For Mandarin, Paraformer is used as the ASR teacher, while for English, SenseVoice Small is adopted. All evaluations are conducted in a simulated online streaming inference setting on a single NVIDIA L20 GPU.

To evaluate the dialogue state control capability of SoulX-Duplug, we construct a modular FD-SDS based on it, leveraging Qwen2.5-7B-Instruct \cite{Yang2024Qwen2.5} as the LLM and IndexTTS-1.5\footnote{\scriptsize\url{https://github.com/Ksuriuri/index-tts-vllm}} \cite{deng2025indextts} as the TTS model. The system is evaluated on selected tasks from the Full-Duplex-Bench (FDB) series, including Pause Handling, Turn Taking, and User Interruption from FDB v1, as well as User Backchannel and User Interruption from FDB v1.5. We further evaluate Freeze-Omni and our system on the Chinese test sets introduced in SoulX-Duplug-Eval to assess cross-lingual performance.
We adopt the official metrics defined in FDB. For Pause Handling, Takeover Rate (TOR) measures the frequency with which the system takes the floor when the user only pauses but has not finished speaking, where a lower value indicates better pause management. For Turn Taking, TOR denotes the proportion of successful turn transitions, and Response Latency (RL) measures the delay between the end of the user’s speech and the start of the system’s response.
For User Interruption v1, TOR evaluates whether the system takes the turn after an interruption, and RL measures the latency of the response following the interruption. For User Backchannel, we report Resume Rate (RsR), defined as the proportion of RESUME behaviors after overlap events. For User Interruption v1.5, we report Respond Rate (RpR), which denotes the proportion of RESPOND behaviors. In addition, Stop Latency (SL) measures the delay between the onset of user interruption and the moment the system stops speaking, and RL measures the latency from the end of the interruption to the system’s subsequent response. We aggregate these metrics to obtain overall indicators for system performance: for overall turn management, we average $1-\text{TOR}$ for Pause Handling with TOR, RsR, and RpR from the other tasks; for overall latency, we compute the mean of all RL and SL values across tasks. ASR is conducted with Paraformer when evaluating on Chinese subsets.
For duplex state prediction, we further evaluate the model on the Easy Turn testset and the English extension introduced in SoulX-Duplug-Eval, conducting further comparison and ablation studies using prediction accuracy (ACC) and inference latency as the evaluation metrics.

\section{Experimental Results}

\begin{table*}[htbp]
\centering
\small
\resizebox{1\linewidth}{!}{
\begin{tabular}{lcccccccccccc}
\toprule
\multirow{5}{*}{\textbf{Lang}} & \multirow{5}{*}{\textbf{Models}} & \multirow{3}{*}{\textbf{Pause Handling}} & \multicolumn{2}{c}{\multirow{3}{*}{\textbf{Turn Taking}}} & \multirow{3}{*}{\textbf{User Backchannel}} & \multicolumn{5}{c}{\textbf{User Interruption}} & \multicolumn{2}{c}{\multirow{3}{*}{\textbf{Overall Score}}} \\
\cmidrule(lr){7-11}
& & & & & & \multicolumn{2}{c}{v1} & \multicolumn{3}{c}{v1.5} \\
\cmidrule(lr){3-3}
\cmidrule(lr){4-5}
\cmidrule(lr){6-6}
\cmidrule(lr){7-8}
\cmidrule(lr){9-11}
\cmidrule(lr){12-13}
& & TOR ($\downarrow$) & TOR ($\uparrow$) & RL ($\downarrow$) & RsR ($\uparrow$) & TOR ($\uparrow$) & RL ($\downarrow$) & RpR ($\uparrow$) & SL ($\downarrow$) & RL ($\downarrow$) & ACC ($\uparrow$) & Latency ($\downarrow$) \\
\midrule
\multirow{9}{*}{EN} & dGSLM & 0.935 & \underline{0.975} & 0.352 & - & 0.917 & 2.531 & - & - & - & 0.653 & 1.442 \\
& PersonaPlex & 0.623 & \textbf{0.992} & \textbf{0.070} & - & \textbf{1.000} & \underline{0.400} & - & - & - & \underline{0.790} & \textbf{0.235} \\
& Moshi & 0.983 & 0.941 & \underline{0.265} & 0.060 & \textbf{1.000} & \textbf{0.257} & 0.500 & 1.160 & 1.470 & 0.504 & 0.788 \\
& Freeze-Omni & 0.562 & 0.336 & 0.953 & 0.800 & 0.867 & 1.409 & 0.720 & 1.420 & \underline{1.350} & 0.632 & 1.283 \\
& Gemini Live & \textbf{0.283} & 0.655 & 1.301 & \underline{0.930} & 0.891 & 1.183 & 0.330 & 2.200 & 2.620 & 0.705 & 1.826 \\
& Sonic & - & - & - & \textbf{0.980} & - & - & 0.240 & 2.250 & 2.750 & 0.610 & 2.500 \\
& GPT-4o & - & - & - & 0.700 & - & - & \textbf{0.780} & \textbf{0.230} & 1.500 & 0.740 & 0.865 \\
\cmidrule(lr){2-13}
& SoulX-Duplug (Ours) & \underline{0.352} & 0.933 & 0.511 & 0.740 & \underline{0.970} & 0.773 & \underline{0.770} & \underline{0.450} & \textbf{1.030} & \textbf{0.812} & \underline{0.691} \\
\midrule
\midrule
\multirow{3}{*}{ZH} & Freeze-Omni & 0.042 & 0.652 & 0.780 & 0.700 & 0.975 & \textbf{0.232} & 0.440 & 1.300 & 1.630 & {0.745} & {0.986} \\
\cmidrule(lr){2-13}
& SoulX-Duplug (Ours) & \textbf{0.038} & \textbf{0.994} & \textbf{0.767} & \textbf{0.800} & \textbf{0.994} & 1.089 & \textbf{0.830} & \textbf{0.380} & \textbf{1.150} & \textbf{0.916} & \textbf{0.847} \\
\bottomrule
\end{tabular}
}
\vspace{2mm}
\caption{Main results on Bilingual Full-Duplex-Bench. For Pause Handling (EN), results are averaged over the Candor and Synthetic subsets. All latency metrics are reported in seconds. The best-performing results are highlighted in \textbf{bold}, and the second-best results are \underline{underlined}.}
\vspace{-3mm}
\label{tab:result-Full-Duplex-Bench}
\end{table*}

\begin{table*}[htbp]
\centering
\small
\resizebox{0.85\linewidth}{!}{
\begin{tabular}{lcccccc}
\toprule
\textbf{Lang} & \textbf{Models} & \textbf{Streaming} & \textbf{ACC$_{Complete}$} ($\uparrow$) & \textbf{ACC$_{Incomplete}$} ($\uparrow$) & \textbf{Latency} ($\downarrow$) & \textbf{Avg. ACC} ($\uparrow$) \\
\midrule
\multirow{4}{*}{EN} & \makecell{SenseVoice En + TEN Turn Detection} & \textcolor{red}{\ding{55}} & 95.60\% & 76.59\% & $latency_{vad}+57$\,ms & {86.10\%} \\
& Smart Turn V2 & \textcolor{red}{\ding{55}} & 83.65\% & 54.85\% & $latency_{vad}+15$\,ms & 69.25\% \\
\cmidrule(lr){2-7}
& SoulX-Duplug (Ours) & \textcolor{green!50!black}{\ding{51}} & 77.67\% & 88.96\% & 240\,ms$^\dag$ & {83.32\%} \\
\midrule
\midrule
\multirow{5}{*}{ZH} & \textcolor{gray}{Easy Turn$^*$} & \textcolor{gray}{\ding{55}} & \textcolor{gray}{96.33\%} & \textcolor{gray}{97.67\%} & \textcolor{gray}{$latency_{vad}+263$\,ms} & \textcolor{gray}{97.00\%} \\
& \makecell{Paraformer + TEN Turn Detection$^*$} & \textcolor{red}{\ding{55}} & 86.67\% & 89.30\% & $latency_{vad}+204$\,ms & 87.99\% \\
& Smart Turn V2$^*$ & \textcolor{red}{\ding{55}} & 78.67\% & 62.00\% & $latency_{vad}+27$\,ms & 70.34\% \\
\cmidrule(lr){2-7}
& SoulX-Duplug (Ours) & \textcolor{green!50!black}{\ding{51}} & 89.33\% & 79.33\% & 240\,ms$^\dag$ & {84.33\%} \\
\bottomrule
\end{tabular}
}
\vspace{2mm}
\caption{Comparison with non-streaming state prediction modules on Bilingual Easy Turn testset. $^*$Results taken from the Easy Turn paper \cite{li2025easy-turn}. Since the Easy Turn testset consists of pre-segmented speech clips, whereas practical deployment requires VAD-based segmentation, the latency of non-streaming models should additionally include $latency_{vad}$. $^\dag$Theoretical latency. Deployment-measured latency are reported in Table~\ref{tab:latency-comparison}.}
\vspace{-6mm}
\label{tab:result-easy-turn}
\end{table*}

\subsection{Results and Analysis}

\subsubsection{Main Results on Bilingual Full-Duplex-Bench}

We construct a cascaded full-duplex dialogue system by integrating SoulX-Duplug as the speech understanding and state management module with Qwen2.5-7B-Instruct \cite{Yang2024Qwen2.5} for response generation and IndexTTS-1.5 \cite{deng2025indextts} for speech synthesis. The complete system is evaluated on Bilingual Full-Duplex-Bench, and the results are summarized in Table~\ref{tab:result-Full-Duplex-Bench}. Overall, the proposed system demonstrates balanced performance across all evaluation dimensions, achieving the highest average score in turn management without exhibiting abnormally low values on any individual metric. Despite adopting a cascaded architecture, it maintains consistently low latency, ranking second-best on average in English and the best in Chinese, which indicates that the streaming turn control module effectively supports real-time interaction.

Considering Turn Taking and Pause Handling jointly, end-to-end continuous-output models (i.e., dGSLM, PersonaPlex, and Moshi) achieve very high turn-taking rates and low response latency. However, these models also exhibit high TOR in Pause Handling, indicating that their strong turn-taking performance is partially achieved by shifting the operating point toward more aggressive response behavior. In contrast, the SoulX-Duplug system attains a TOR of 0.933 in Turn Taking, which is close to the best reported performance, while maintaining a substantially lower probability of inappropriate interruption in Pause Handling than dGSLM, PersonaPlex, and Moshi. 
Moreover, our system consistently outperforms Freeze-Omni across all three related metrics in both Chinese and English test sets. Compared with Gemini, our system improves the Turn Taking TOR by more than 40\%. Although Gemini achieves a slightly lower TOR in Pause Handling, its response latency is more than twice that of SoulX-Duplug, reflecting a different trade-off between conservativeness and responsiveness.

In the User Interruption scenario, the v1 and v1.5 settings share the same test data but adopt different evaluation protocols. SoulX-Duplug achieves strong performance in both the v1 TOR and the v1.5 Respond Rate, ranking the highest among Chinese models and second highest in English. In terms of latency, it achieves the lowest response latency under the v1.5 metric, and its stop latency is second only to GPT-4o in English. An interesting observation is that Freeze-Omni exhibits a high TOR under the Chinese v1 metric, while its Respond Rate is only 0.44 and the proportion of Resume cases reaches 0.47 under v1.5 evaluation. Manual inspection of a subset of audio samples shows that the model occasionally fails to respond to user interruption. The discrepancy suggests that it may mistakenly capture the latter portion of the model’s first response under the evaluation protocol of v1, leading to inflated TOR and latency scores. Similar metric patterns are observed for Moshi and Gemini on the English test set.

For User Backchannel handling, models with high Resume Rates tend to exhibit longer stop latency under interruption test, often exceeding 1\,s and in some cases 2\,s. In contrast, GPT-4o achieves a stop latency of 0.23\,s but with a Resume Rate of 0.7, indicating a clear trade-off between backchannel handling and responsiveness to user interruption. SoulX-Duplug attains a stop latency of 0.45\,s with a Resume Rate of 0.74, representing a balanced point without extreme bias toward either objective.

\subsubsection{Comparison Against Non-Streaming State Prediction Modules}

As presented in Table~\ref{tab:result-easy-turn}, we evaluate the streaming SoulX-Duplug against open-source non-streaming modules on the Bilingual Easy Turn testset, comparing their performance on latency and dialogue state prediction. Despite streaming modeling, under a fair comparison with a pipeline system composed of a state-of-the-art ASR model and a 7B-parameter TEN Turn Detection, where neither module is trained on the Easy Turn train set, SoulX-Duplug’s accuracy is only about 3\% lower, which is within an acceptable range.
This result indicates that the unified text-guided streaming formulation does not compromise turn detection accuracy, even when compared with large-scale non-streaming baselines.

More importantly, the streaming design of SoulX-Duplug enables consistently low and stable latency. In contrast, in practical deployment scenarios, non-streaming approaches have to rely on an external VAD model to segment audio before inference. Such VAD-based truncation introduces additional delay, which is often several hundred milliseconds (e.g., the latency of VAD model used for comparison is reported as 500\,ms in Flexduo \cite{liao2025flexduo}). Moreover, this latency may increase with longer input. As a result, although non-streaming systems may appear to achieve higher accuracy, they incur greater and less predictable response delays in real-world applications. 

\subsubsection{Comparison Across Streaming State Prediction Modules}

\begin{table}[H]
\centering
\resizebox{\linewidth}{!}{
\begin{tabular}{lcccc}
\toprule
& \multicolumn{2}{c}{\textbf{SoulX-Duplug (Ours)}} & \multirow{3}{*}{\textbf{FlexDuo}$^\dag$} & \multirow{3}{*}{\textbf{VAD}$^\dag$} \\
\cmidrule(lr){2-3}
& EN & ZH & & \\
\midrule
\textbf{Latency} ($\downarrow$) & \textbf{205\,ms} & \textbf{295\,ms} & 343\,ms & 500\,ms \\
\bottomrule
\end{tabular}
}
\vspace{2mm}
\caption{Latency comparison of different streaming state prediction modules. $^\dag$Results reported in the FlexDuo paper.}
\vspace{-5mm}
\label{tab:latency-comparison}
\end{table}

To quantify the standalone latency of SoulX-Duplug within a practical dialogue system, we conduct a measurement on the Turn Taking task of Full-Duplex-Bench. Specifically, we record the first-packet latency of the downstream LLM and TTS modules during inference. The latency of SoulX-Duplug is then estimated by subtracting the LLM and TTS first-packet latency from the reported speak latency in the evaluation metric. The results are concluded in Table~\ref{tab:latency-comparison}.

Since the model and evaluation set of FlexDuo \cite{liao2025flexduo} are not publicly available, it is not possible to assess its latency under the identical experimental setting. Nevertheless, for streaming modules, latency typically varies only slightly across test sets, so we take the numbers from FlexDuo paper for comparison. As shown in the table, among streaming state prediction modules, SoulX-Duplug achieves an average latency of 250\,ms across both English and Chinese in practical deployment, substantially lower than both FlexDuo and VAD-based approaches. This improvement can be attributed to the small chunk size adopted in SoulX-Duplug and its lightweight components, especially the backbone LLM. Notably, the practical latency (250\,ms) is close to the theoretical latency (240\,ms), confirming that teacher-forced inference does not introduce additional delay.

\subsection{Ablation Study}

\begin{table}[H]
\centering
\small
\resizebox{\linewidth}{!}{
\begin{tabular}{lccc}
\toprule
\textbf{Models} & \textbf{ACC$_{Com}$} ($\uparrow$) & \textbf{ACC$_{Incom}$} ($\uparrow$) & \textbf{Avg.} ($\uparrow$) \\
\midrule
SoulX-Duplug (Proposed) & \textbf{89.33\%} & \textbf{79.33\%} & \textbf{84.33\%} \\
\hspace{1em}w/o ASR Pretraining & 83.00\% & 78.00\% & 80.50\% \\
\hspace{1em}w/o Teacher-Forced Inference & 78.33\% & 68.67\% & 73.50\% \\
\bottomrule
\end{tabular}
}
\vspace{2mm}
\caption{Ablation results on the Easy Turn testset (Zh).}
\vspace{-5mm}
\label{tab:easyturn-ablation}
\end{table}

We conduct ablation experiments to examine the effects of the training and inference strategies used in SoulX-Duplug. The results are demonstrated in Table~\ref{tab:easyturn-ablation}. 

First, when the first and second stages of ASR pretraining are removed, the state prediction accuracy decreases noticeably. This observation indicates that the model relies on its ASR capability when performing the state prediction task. The ASR pretraining stage provides improved semantic representations, which in turn benefits the third-stage supervised fine-tuning and downstream inference.


In addition, when external-guided ASR is removed during inference, the prediction accuracy degrades. This result validates the effectiveness of our Hybrid E2E Training with Teacher-Forced Inference strategy. Under the extremely short chunk setting, externally provided ASR outputs offer more reliable textual supervision, which stabilizes semantic interpretation. The performance drop further confirms that the model indeed leverages explicit textual semantic information to support streaming state prediction.

\subsection{Further Discussion}

Throughout the course of this research, we developed several observations and reflections regarding the design and deployment of FD-SDSs:

\begin{itemize}
    \item Streaming ASR with very small chunk sizes remains inherently challenging. When the chunk duration is short, acoustic segments frequently cut across phoneme, syllable, or word boundaries. This issue is particularly pronounced in English, where words can be easily fragmented across adjacent chunks, leading to recognition instability and transient errors. As a result, a certain degree of prediction fluctuation is unavoidable under strict low-latency constraints. Moreover, in real-time streaming settings, the strong contextual modeling capacity of LLM-based approaches are constrained by incremental decoding and limited future context. Therefore, LLM-based ASR models do not necessarily outperform conventional architectures such as RNN-T in terms of the balance between decoding speed and recognition accuracy.
    \item Although recent end-to-end FD-SDMs have demonstrated strong empirical performance and considerable potential, they typically require large-scale training data and substantial computational resources. In contrast, modular systems are comparatively easier to implement and maintain. When performance issues arise, individual components can be adjusted or replaced without retraining the entire system. This flexibility may make modular designs more suitable for practical deployment \cite{liu2025xtalk}.
    \item Finally, current research efforts provide relatively limited support for real-time applications. There remains a need for more mature and accessible open-source streaming speech encoders and ASR models. Continued development in this direction would facilitate the advances of truly practical FD-SDSs.
\end{itemize}

\section{Conclusion}

In this work, we introduce SoulX-Duplug, a pluggable streaming state prediction module designed for real-time full-duplex speech conversation. 
Through text-guided streaming state prediction and hybrid 3-stage training with teacher-forced inference, SoulX-Duplug enables low-latency semantic-aware streaming dialogue management. On Full-Duplex-Bench, the SoulX-Duplug-based system outperforms existing models with the best overall performance in turn management and latency-related metrics. Additional experiments further validate the effectiveness of the proposed approaches.
Moreover, we release supplementary evaluation sets for Easy Turn testset and Full-Duplex-Bench, referred to as SoulX-Duplug-Eval, to support more comparable cross-lingual benchmarking. 
Both SoulX-Duplug and SoulX-Duplug-Eval have been open-sourced. We hope that these resources will contribute to continued progress in real-time spoken dialogue modeling.

\section{Generative AI Use Disclosure}

We used generative AI tools solely for language polishing of the manuscript. No generative AI tools were used for drafting, generating, or modifying the scientific content, experimental design, results, or conclusions of this work.

\bibliographystyle{IEEEtran}
\bibliography{mybib}

\end{document}